УДК 62.50

**Метод поиска функции Ляпунова для анализа устойчивости нелинейных систем с использованием генетического алгоритма**

А.М. Зенкин, А.А. Перегудин, А.А. Бобцов

**Аннотация**

**Введение.** В статье рассматривается широкий класс гладких непрерывных динамических нелинейных систем (объектов управления) с измеряемым вектором состояния. Ставится задача поиска специальной функции (функции Ляпунова), которая в рамках второго метода Ляпунова гарантирует асимптотическую устойчивость для описанного выше класса нелинейных систем. Хорошо известно, что поиск функции Ляпунова является «краеугольным камнем» математической теории устойчивости. Методы подбора или поиска функции Ляпунова для анализа устойчивости замкнутых линейных стационарных систем хорошо изучены, также, как и для нелинейных объектов с явно выраженными линейной динамической и нелинейной статической частями (см. работы Лурье, Якубовича, Попова и многих других). Однако универсальных подходов к поиску функции Ляпунова для более общего класса нелинейных систем так и не выявлено. Существует большое разнообразие методов поиска функции Ляпунова для нелинейных систем, но все они работают в рамках заданных на структуру объекта управления ограничений. В этой статье предлагается ещё один подход, который позволяет дать специалистам в области теории автоматического управления новый инструмент/механизм поиска функции Ляпунова для анализа устойчивости гладких непрерывных динамических нелинейных систем с измеряемым вектором состояния. Суть предлагаемого подхода заключается в представлении некоторой функции через сумму нелинейных слагаемых, представляющих собой возведенные в положительные степени элементы вектора состояния объекта, умноженные на неизвестные коэффициенты. Далее с использованием генетического алгоритма осуществляется подбор неизвестных коэффициентов, которые должны обеспечить указанной функции все необходимые условия для функции Ляпунова (в рамках второго метода Ляпунова). **Метод.** Предложен новый подход поиска функции Ляпунова для анализа устойчивости нелинейных объектов через разложение неизвестной функции Ляпунова в сумму нелинейных слагаемых с неизвестными коэффициентами, поиск которых осуществляется классическим генетическим алгоритмом, включающим операции мутации, селекции и кроссовера. В отличии от большинства предлагаемых методов близкой идеологии подход на основе генетического алгоритма не требует обучающей выборки, которая накладывает ограничения в виде структуры объектов управления, включающихся в нее. Потенциальная функция Ляпунова формируются итеративно путем применения генетических операций к произвольно заданной начальной популяции коэффициентов. **Основные результаты.** Эффективность предложенного метода продемонстрирована на примере математической модели маятника путем варьирования диапазона коэффициентов и размера области, в которой рассматривается система. Также были проведены компьютерные симуляции с фиксированным количеством итераций и изменяющимся размером популяции. По результатам данных компьютерных симуляций была установлена зависимость количества успешно найденных функций Ляпунова от количества итераций генетического алгоритма. Было проведено сравнение предложенного метода с аналогичным подходом, использующим при построении функции Ляпунова близкую идеологию. **Обсуждение.** В работе был предложен новый подход поиска функции Ляпунова для анализа устойчивости гладких непрерывных динамических нелинейных систем с измеряемым вектором состояния. Основная идея подхода заключается в подборе с использованием генетических алгоритмов некоторых неизвестными коэффициентов/параметров, которые представляют собой множители для известных нелинейных функций в совокупности, представляющих собой функцию Ляпунова. Предложенный в данной работе метод обеспечивает поиск функции Ляпунова и по быстродействию превосходит известные аналоги. Предложенное решение отличается от известного более общего подхода, поскольку рассматривается разложение потенциальной функции Ляпунова в ряд Тейлора с неизвестными коэффициентами, вместо использования контрпримеров или шаблонных функций Ляпунова.

**Ключевые слова:** функция Ляпунова, математическая устойчивость, машинное обучение, генетический алгоритм, функция ценности, математический маятник.

# Lyapunov function search method for analysis of nonlinear systems stability using genetic algorithm

A.M. Zenkin, A.A. Peregudin, A.A. Bobtsov

**Abstract**
**Introduction**. The paper considers a wide class of smooth continuous dynamic nonlinear systems (control objects) with a measurable vector of state. The task is to find a special function (Lyapunov function), which in the framework of the second Lyapunov method guarantees asymptotic stability for the above-described class of nonlinear systems. It is well known that the search for a Lyapunov function is the "cornerstone" of mathematical stability theory. Methods of selection or search for the Lyapunov function for stability analysis of closed linear stationary systems are well studied, as well as for nonlinear objects with explicit linear dynamic and nonlinear static parts (see works by Lurie, Yakubovich, Popov and many others). However, no universal approaches to the search for the Lyapunov function for a more general class of nonlinear systems have been identified. There is a large variety of methods for searching the Lyapunov function for nonlinear systems, but all of them operate within constraints set on the control object structure. In this article we propose another approach, which allows to give specialists in the field of automatic control theory a new tool/mechanism of search for Lyapunov function for stability analysis of smooth continuous dynamic nonlinear systems with measurable state vector. The essence of the proposed approach consists in representation of some function through sum of nonlinear terms, representing the elements of the object state vector multiplied by unknown coefficients, raised in positive degrees. Then, using genetic algorithm, unknown coefficients are selected, which should provide the function with all necessary conditions for Lyapunov function (in the framework of the second Lyapunov method). **Method**. A new approach of Lyapunov function search for stability analysis of nonlinear objects through decomposition of unknown Lyapunov function into a sum of nonlinear terms with unknown coefficients, which search is performed by classical genetic algorithm including mutation, selection, and crossover operations. Unlike most proposed methods of close ideology, the genetic algorithm approach does not require a training sample, which imposes restrictions in the form of the structure of control objects included in it. The potential Lyapunov function is formed iteratively by applying genetic operations to an arbitrary initial population of coefficients. **Main Results**. The effectiveness of the proposed method has been demonstrated on the example of a mathematical model of a pendulum by varying the range of coefficients and the size of the domain in which the system is considered. Computer simulations with a fixed number of iterations and varying population size were also performed. Based on the results of these computer simulations, the dependence of the number of successfully found Lyapunov functions on the number of iterations of the genetic algorithm was established. The proposed method was compared with a similar approach that uses a similar ideology for the Lyapunov function. **Discussion**. In this paper a new approach of Lyapunov function search for stability analysis of smooth continuous dynamical nonlinear systems with measurable vector of state has been proposed. The main idea of the approach is to select, using genetic algorithms, some unknown coefficients/parameters, which are multipliers for known nonlinear functions in aggregate, representing the Lyapunov function. The method proposed in this paper provides a Lyapunov function search and exceeds the known analogues in terms of speed. The proposed solution differs from the known more general approach because it considers the expansion of the potential Lyapunov function into a Taylor series with unknown coefficients, instead of using counterexamples or template Lyapunov functions.

**Keywords:** Lyapunov function, mathematical stability, machine learning, genetic algorithm, value function, mathematical pendulum.

**Acknowledgements**
The study was supported by a grant from the Russian Science Foundation (Project 23-16-00224).

## Введение

Анализ устойчивости нелинейных динамических систем, описываемых дифференциальными уравнениями, является фундаментальной проблемой в теории управления. Для анализа устойчивости часто используется второй метод Ляпунова, который заключается в поиске положительной функции, убывающей вдоль траекторий динамической системы [1–3]. Поиск функции Ляпунова в общем виде является трудной и нетривиальной задачей, которая активно исследуется научным сообществом. Существуют аналитические методы поиска функции Ляпунова для нелинейных систем [4–6], но они имеют высокую вычислительную сложность и не всегда позволят подобрать функцию Ляпунова [7]. Наряду с этим существуют численные методы поиска [8], которые также достаточно сложны и в некоторых случаях не имеют аналитической обоснованности [9]. Однако для нелинейных систем не существует общего метода поиска функции Ляпунова [10]. Существуют подходы с использованием технологий машинного обучения, в которых вручную подбираются шаблоны функций Ляпунова, на основе которых осуществляется обучение алгоритма и дальнейший поиск [11,12]. Наряду с зарубежными исследователями аналогичной тематикой активно занимаются отечественные ученые. Так, в работе [13] рассматривается метод по восстановлению функциональных зависимостей, используя статистические данные. Метод основан на статистической выборке, базисных функциях и статистических показателях. для восстановления функциональных зависимостей по статистическим данным. Авторами [14] предложен метод формирования функции Ляпунова с использованием нейронных сетей и целочисленного линейного программирования.

Однако универсальных подходов к поиску функции Ляпунова для более общего класса нелинейных систем так и не выявлено. Существует большое разнообразие методов поиска функции Ляпунова для нелинейных систем, но все они работают в рамках заданных на структуру объекта управления ограничений. В данной статье предлагается развитие данных подходов, используя вместо подбираемых вручную шаблонов функцию, представленную в виде ряда Тейлора с неизвестными коэффициентами/параметрами, поиск которых осуществляется при помощи машинного обучения. Данная работа расширяет результаты, полученные авторами [15], в части улучшения ряда качественных характеристик, связанных с синтезом функции Ляпунова.

В статье представлен метод синтеза (поиска) функции Ляпунова для нелинейных систем в общем виде. Метод основан на разложении потенциальной функции Ляпунова в бесконечную сумму степенных функций и эффективный алгоритм машинного обучения, осуществляющий роль поиска коэффициентов ряда. В качестве алгоритма машинного обучения используется генетический алгоритм из семейства метаэвристических алгоритмов.

Эффективность предложенного метода была оценена на двух нелинейных системах. Были рассмотрены самые широкие области, для каждой из которых была найдена функция Ляпунова. Предложенный метод достиг более широких областей выполнения условий теоремы Ляпунова, чем подход, представленный в работе [15], за более короткое время.

Необходимо отметить, что предложенный в этой статье метод дает самые сильные гарантии в рамках альтернатив (асимптотическая устойчивость) и не полагается на заранее подготовленные вручную шаблоны функций Ляпунова.

В целом, в работе представлен новый метод синтеза функции Ляпунова, который 1) не требует необходимости подбора шаблонов функций вручную 2) работает быстрее и охватывает более широкую область выполнения условий теоремы Ляпунова, чем другие современные инструменты, в частности метод, предложенный в работе [15].

**Постановка задачи**

Рассмотрим нелинейную автономную систему (1):

$$\dot{x}(t) = f\big(x(t)\big), \tag{1}$$

где $x(t) \in \mathbb{R}^n$ – состояния в момент времени $t$, $f \colon D \to \mathbb{R}^n$ – липшицево отображение из произвольной области $\mathrm{D} \in \mathbb{R}^n$. Предположим, что в области $D$ имеется точка/положение равновесия, то есть такая точка $\bar{x} \in D$, для которой выполнено $f(\bar{x}) = 0$.

**Замечание 1.** Следует отметить, что анализ устойчивости положения равновесия $\bar{x}$ системы (1) является следствием решения задачи синтеза стабилизирующего регулятора. Иными словами, математическая модель (1) может рассматриваться в качестве замкнутой системы, полученной с использованием обратной связи для некоторого номинального нелинейного объекта вида

$$\dot{x}(t) = f_0\big(x(t)\big) + g(x, u), \tag{2}$$

где функция $u(x, t)$ – сигнал управления, обеспечивающий замкнутой системе асимптотическую устойчивость и выбираемый таким образом, чтобы для (1) существовала функция Ляпунова $V(x)$ с заданными свойствами (то есть $V(x) > 0$ и $\dot{V}(x) < 0$ во всех точках, кроме точки равновесия).

**Допущение 1.** Точка равновесия $\bar{x}$ замкнутой системы (1) асимптотически устойчива, и существует функция Ляпунова $V(x)$ такая, что $V(\bar{x}) = 0, V(x) > 0$ и $\dot{V}(x) < 0$ при $x \neq \bar{x}$.

**Допущение 2.** Функция $V(x)$ допускает разложение в ряд Тейлора в окрестности положения равновесия $\bar{x}$, сходящийся к ней во всех точках $x \in D$.

Согласно допущениям 1–2 для системы (1) существует функция Ляпунова, представимая в виде ряда Тейлора

$$V(x_1, \dots, x_n) = \sum_{m=0}^{N} \sum_{k_1 + \dots + k_n = m} \prod_{i=1}^{n} \frac{(x_i - \bar{x})^{k_i}}{k_i!} \frac{\partial^m V}{\partial x_1 \dots \partial x_n}(\bar{x}_1, \dots, \bar{x}_2) + R_N(x_1, \dots, x_n), \tag{3}$$

где $R_N$ – остаточный член ряда Тейлора.

Назовем *потенциальной функцией Ляпунова* функцию вида

$$L(x_1, \dots, x_n) = \sum_{m=0}^{N} \sum_{k_1 + \dots + k_n = m} \prod_{i=1}^{n} (x_i - \bar{x})^{k_i} \, p_{k_1, \dots, k_n}, \tag{4}$$

где $p_{k_1, \dots, k_n} \in \mathbb{R}$ – числовые параметры, подлежащие определению. Если эти параметры найдены точно, то есть, если выполнено

$$p_{k_1, \dots, k_n} = \frac{1}{k_i!} \frac{\partial^m V}{\partial x_1 \dots \partial x_n}(\bar{x}_1, \dots, \bar{x}_2),$$

то потенциальная функция (4) совпадает с функцией Ляпунова (3) с точностью до остаточного члена. Из допущения 2 следует, что $\lim\limits_{N \to \infty} R_N(x_1, \dots, x_n) = 0$ при всех $(x_1, \dots, x_n) = x \in D$, следовательно в области $D$ функция $V$ может быть сколько угодно хорошо приближена функцией $L$ за счет правильного выбора параметров $p_{k_1, \dots, k_n}$.

Таким образом с учетом Допущений 1–2 и формулы (4) сформулируем дополнительную цель данной статьи, как разработку нового метода поиска параметров $p_{k_1,\ldots,k_n}$, обеспечивающих (4) свойства функции Ляпунова.

Поиск параметров $p_{k_1,\ldots,k_n}$ предполагается осуществлять на базе генетического алгоритма. Работа генетического алгоритма будет построена на предположении, что если параметры $p_{k_1,\ldots,k_n}$ выбраны правильно, то для функции $L$ выполняются условия теоремы Ляпунова:

$$L(x_1, x_2, \ldots, x_n) > 0, \tag{5}$$

$$\dot{L}(x_1, x_2, \ldots, x_n) = \sum_{i=0}^{n} \frac{\partial L}{\partial x_i} \dot{x}_i = \begin{bmatrix} \frac{\partial L}{\partial x_1} & \frac{\partial L}{\partial x_2} & \ldots & \frac{\partial L}{\partial x_n} \end{bmatrix} \begin{bmatrix} f_1(x) \\ f_2(x) \\ \vdots \\ f_n(x) \end{bmatrix} < 0. \tag{6}$$

По определению $L(\bar{x}) = 0$, поэтому данное условие не будет проверяться в процессе поиска функции. Оценка выполнения условий (5) и (6) будет производиться с помощью функции ценности (функции затрат) специального вида.

**Формирование функции ценности для генетического алгоритма**

Генетический алгоритм [16,17] относится к метаэвристическим алгоритмам, в основе которых лежит биологический принцип естественного отбора, основанный на соревновании между популяциями. Оценкой эффективности популяции служит целевая функция, которая показывает, насколько хорошо данный закон ее оптимизирует, также ее называют функцией затрат. Комбинация параметров, которые наиболее успешно справились со своей задачей, порождают новое поколение, которое должно быть эффективнее предыдущего. Задачей является выявить наиболее эффективные наборы комбинаций из текущего поколения и передать их в следующее с использованием генетических операций для порождения более эффективных комбинаций.

Обозначим функцию ценности, как $J(p)$, где p – набор коэффициентов вида $p_{k_1,\ldots,k_n}$, которые подлежат оптимизации (в данной работе они являются коэффициентами перед каждым членом ряда Тейлора). Данная функция принимает значения в диапазоне от 0 до 1 включительно, где 0 – наилучшее решение, 1 – наихудшее решение. Пусть имеется нелинейная автономная система (1) и потенциальная функция Ляпунова (4), рассмотрим ограниченную область $\Omega$, в каждой точки которой будем проверять условия теоремы Ляпунова. Обозначим точки, которые удовлетворяют условиям теоремы, как X, не удовлетворяют – Y.

$$\nabla L(x_1, x_2, \ldots, x_n) f(x_1, x_2, \ldots, x_n) = \sum_{i=1}^{n} \frac{\partial L}{\partial x_i} \dot{x}_i = \begin{bmatrix} \frac{\partial L}{\partial x_1} & \frac{\partial L}{\partial x_2} & \ldots & \frac{\partial L}{\partial x_n} \end{bmatrix} \begin{bmatrix} f_1(x_1, x_2, \ldots, x_n) \\ f_2(x_1, x_2, \ldots, x_n) \\ \vdots \\ f_n(x_1, x_2, \ldots, x_n) \end{bmatrix} \tag{7}$$

$$X = \{x \in \mathbb{R}^n | \nabla L(x_1, x_2, \ldots, x_n) \cdot f(x_1, x_2, \ldots, x_n) < 0 \wedge L(x_1, x_2, \ldots, x_n) > 0\} \tag{8}$$

$$Y = \{x \in \mathbb{R}^n | \nabla L(x_1, x_2, \ldots, x_n) \cdot f(x_1, x_2, \ldots, x_n) \geq 0 \vee L(x_1, x_2, \ldots, x_n) \geq 0\} \tag{9}$$

Тогда эффективность параметров, подобранных алгоритмом, будет определяться, как отношение числа точек, которые не удовлетворяют (9) теореме Ляпунова к общему числу точек:

$$J(p) = \frac{|Y|}{|X| + |Y|}, \tag{10}$$

где $0 \leq J \leq 1$. Таким образом при $J = 0$ будут выполнены все условия теоремы Ляпунова.

**Алгоритм 1: Поиск потенциальной функции Ляпунова**

формирование начального поколения из N наборов параметров p
установка счетчика итерации k = 0
определение вероятности размножения $P_c$
определение вероятности мутации $P_m$
**начало цикла (нц)**
**пока**
| t < $t_{MAX}$
**выполнять**
    размножение наборов параметров p с вероятностью $P_c$
    мутация наборов параметров p с вероятностью $P_m$
    **начало цикла (нц)**
    **для**
    i от 1 до N с шагом 1
    **повторять**
        Вычислить производную функцию Ляпунова $L(x_1, x_2, ..., x_n)$ по формуле (6)
        Вычислить функцию Ляпунова $L(x_1, x_2, ..., x_n)$
        **Если** $\nabla L(x_1, x_2, ..., x_n) \cdot f(x_1, x_2, ..., x_n) < 0 \wedge L(x_1, x_2, ..., x_n) > 0$
            **то** увеличить $X$ на один
            **иначе** увеличить $Y$ на один
        **конец если**
        рассчитать функцию ценности J по формуле (10)
    **конец цикла (кц)**
    формирование нового поколения (селекция) на основе J
**конец цикла (кц)**
вернуть набор параметров p с минимальной функцией ценности J
**Если** J > 0
    **то** решение не найдено
    **иначе** решение найдено

Алгоритм 1 работает следующим образом: создается начальная популяция потенциальных функций Ляпунова (размер популяции выбирается на стадии применения алгоритма исходя из имеющихся вычислительных ресурсов), затем они оцениваются на основе заданных критериев, и лучшие функции сохраняются. Затем происходит скрещивание и мутация функций, чтобы создать новые «потомки», которые также оцениваются и сохраняются, если они лучше предыдущих. Этот процесс повторяется до тех пор, пока не будет найдена наилучшая функция Ляпунова. В качестве обратной связи для алгоритма служит предложенная функция затрат или как ее еще называют функция приспособленности, в основе которой лежит проверка выполнения условий теоремы Ляпунова. Графическая визуализация предложенного алгоритма представлена на рисунке 1.

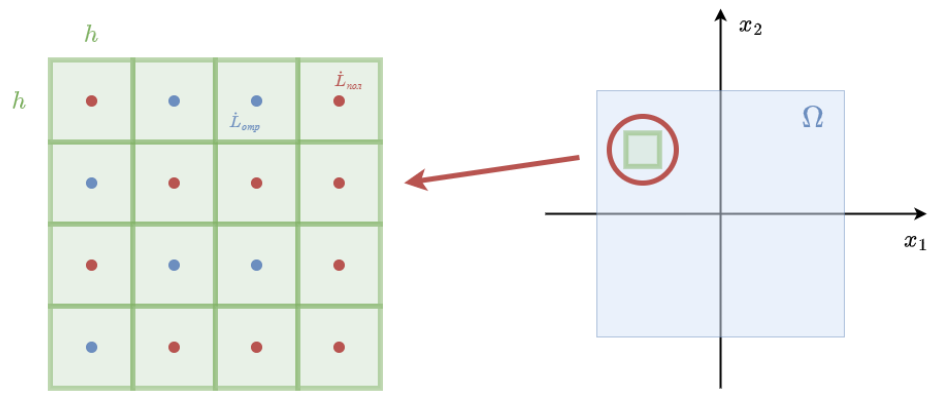

Рис. 1 – Визуализация алгоритма работы функции ценности для двух состояний системы
Fig. 1 – Visualization of the value function algorithm

**Пример реализации алгоритма**

Для иллюстрации работоспособности предложенного метода рассмотрим пример поиска функции Ляпунова для нелинейных систем с применением генетического алгоритма. В качестве объекта исследования был взят простейший математический маятник вида:

$$\frac{d}{dt}\begin{bmatrix} x_1 \\ x_2 \end{bmatrix} = \begin{bmatrix} x_2 \\ -\frac{g}{l} \cdot \sin(x_1) - b \cdot x_2 \end{bmatrix}, x = \begin{bmatrix} \theta \\ \dot{\theta} \end{bmatrix}, \tag{11}$$

$$\left\{ \begin{bmatrix} x_1 \\ x_2 \end{bmatrix} \quad \begin{matrix} -\pi < x_1 < \pi \\ |x_2| < \pi \end{matrix} \right\}, \bar{x} = (0; 0), b \geq 0, \tag{12}$$

где $\bar{x}$ – точка равновесия системы, соответствующая нижнему положению маятника; $\theta$ – угловое отклонение маятника от положения равновесия; $\dot{\theta}$ – угловая скорость маятника; $g$ – ускорение свободного падения; $l$ – длина маятника; $b$ – коэффициент трения. Примем $\frac{g}{l} = 1$, $b = 1$.

Представим потенциальную функцию Ляпунова по формуле (4), разложенную до третьей степени ($n = 2, N = 3$), а именно:

$$L(x_1, x_2) = p_{1,0}x_1 + p_{0,1}x_2 + p_{2,0}x_1^2 + p_{1,1}x_1x_2 + p_{0,2}x_2^2 +$$
$$+ p_{3,0}x_1^3 + p_{2,1}x_1^2x_2 + p_{1,2}x_1x_2^2 + p_{0,3}x_2^3 \tag{13}$$

В качестве начальных условий для генетического алгоритма были взяты параметры, указанные в таблице 1.

Таблица 1 – Начальные условия генетического алгоритма для поиска функции Ляпунова системы (11)-(12)

| Параметр | Значение |
|---|---|
| Диапазон коэффициентов | [-2;2] |
| Область $\Omega$ | 1,0×1,0 |
| Размер популяции | 1000 |
| Вероятность мутации | 20% |
| Вероятность кроссинговера | 40% |
| Процент лучших наборов параметров, переходящих в следующую популяцию | 1% |

Результатом работы алгоритма стала функция

$$L(x_1, x_2) = 8x_1^2 + 8x_1x_2 + 9x_2^2 - x_1^3 + 3x_1^2x_2 - x_2^3, \tag{14}$$

удовлетворяющая условиям (6) и (7) на выбранной области $\Omega$. Процесс эволюции параметров для нелинейной системы (11)-(12) с использованием предложенного метода показан на рисунке 2.

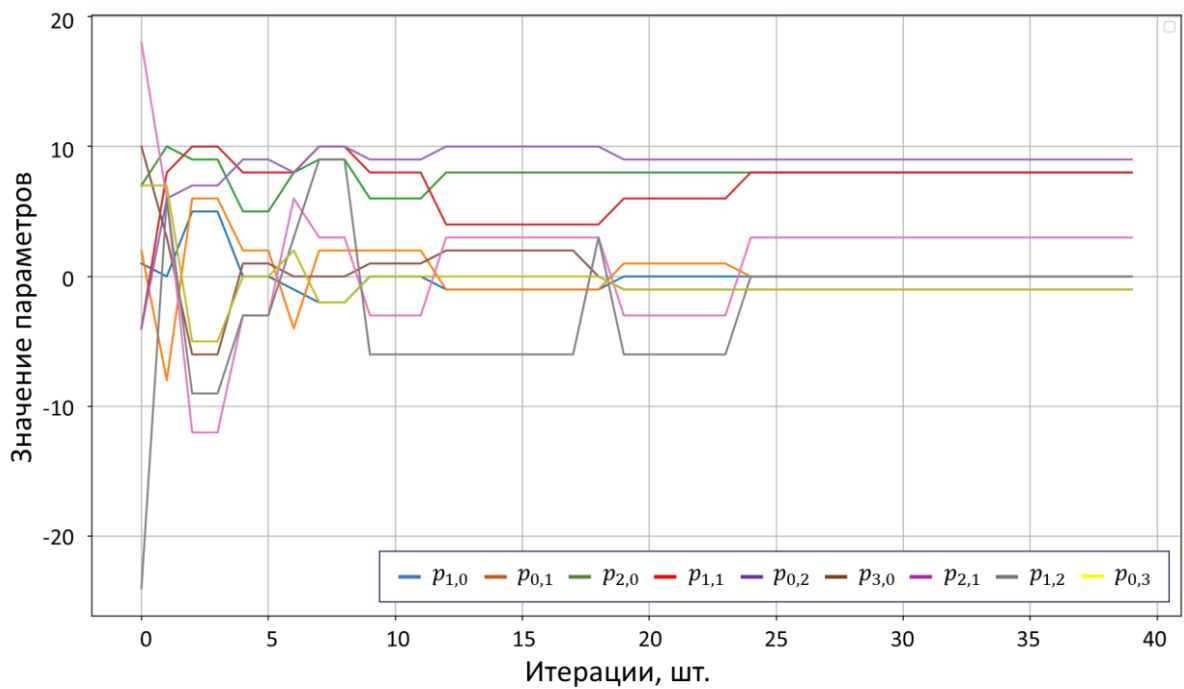

Рис. 2 – Процесс эволюции параметров функции Ляпунова для нелинейной системы (11)-(12) с начальными параметрами, указанными в таблице 1

Fig. 2 - Process of evolution of the Lyapunov function parameters for the nonlinear system (11)-(12) with the initial parameters shown in Table 1

На рисунке 3 показана эволюция функции $\dot{L}(x_1, x_2)$ с шагом в 10 итераций. В начале формируется популяция из 1000 наборов параметров, подобранных случайным образом в диапазоне [-2;2], то есть 1000 различных потенциальных функций Ляпунова, каждой из которых присваивается значение, отражающее ее эффективность с помощью предложенной функции затрат (10). Далее отбираются наиболее эффективные функции и передаются в следующую итерацию с использованием генетических операций для порождения более эффективных функций Ляпунова.

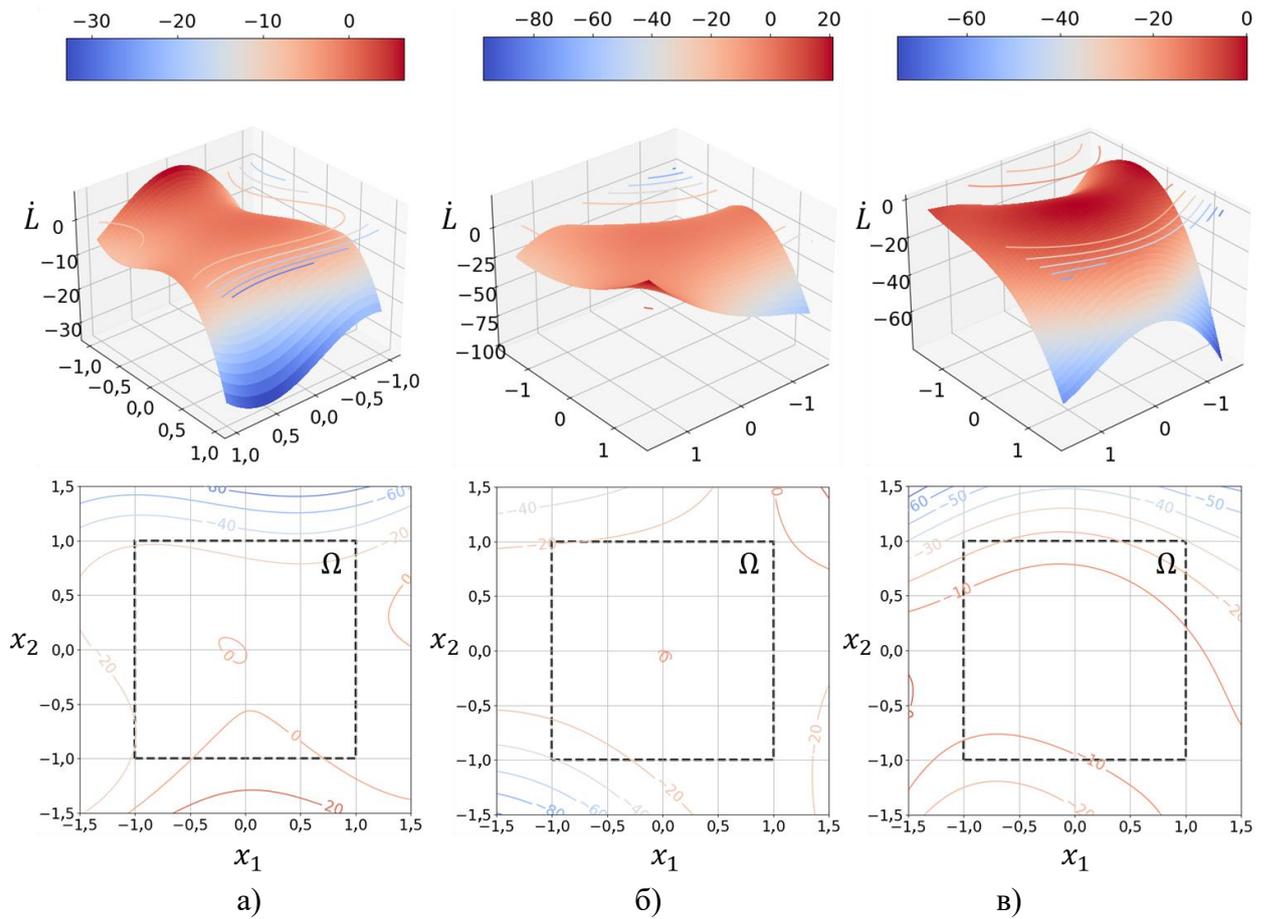

Рис. 3 – Визуализация процесса эволюции потенциальной функции Ляпунова с использованием предложенного метода а) десятое поколение (итерация); б) двадцатое поколение (итерация); в) тридцатое поколение (итерация)

Fig. 3 – Visualization of the evolution process of the potential Lyapunov function using the proposed method a) tenth generation (iteration); b) twentieth generation (iteration); c) thirtieth generation (iteration)

Эффективность предложенного метода была проверена, путем варьирования диапазона коэффициентов $\Delta d$ [-1;1], [-2;2], [-3;3], [-4;4], [-5;5], [-6;6], [-7;7], [-8;8], [-9;9], [-10;10] и [-20;20], перед многочленами потенциальной функции Ляпунова, а также размера области $\Omega$ 0,1×0,1, 0,5×0,5 и 1,0×1,0. Результаты производительности, выраженных временем на поиск функции Ляпунова представлены в таблице 2.

Таблица 2 – Производительность предложенного метода, выраженные временем (t, сек), в зависимости от диапазона коэффициентов $\Delta d$ и размера области $\Omega$

| $\Omega$ | диапазон $\Delta d$ | | | | | | | | | |
|---|---|---|---|---|---|---|---|---|---|---|
| | [-1;1] | [-2;2] | [-3;3] | [-4;4] | [-5;5] | [-6;6] | [-7;7] | [-8;8] | [-9;9] | [-20;20] |
| 0,1×0,1 | 1,43 | 1,45 | 1,85 | 2,39 | 2,92 | 2,82 | 4,36 | 3,14 | 3,27 | 5,41 |
| 0,5×0,5 | - | 4,04 | 3,78 | 3,87 | 4,59 | 4,07 | 5,69 | 5,22 | 7,28 | 13,75 |
| 1,0×1,0 | - | 7,22 | 7,6 | 10,13 | 10,01 | 14,07 | 13,59 | 15,46 | 12,83 | 18,27 |

Как можно видеть, увеличение области усложняет поиск функции Ляпунова, что является достаточно логичным, так как на это уходит больше вычислительных ресурсов. Однако из полученных результатов можно сделать вывод, что для синтеза системы (11)-(12) лучше всего подходит диапазон коэффициентов [-2;2], так как имеет наименьшие вычислительные затраты, выраженные временем. Также отметим зависимость времени поиска функции Ляпунова от диапазона коэффициентов $\Delta d$. Диапазон [-1;1] может не обеспечивать необходимую гибкость алгоритма для поиска функции Ляпунова.

Эффективность методов проверена путем варьирования размеров популяции от 10 до 70 с шагом 10 с фиксированным количеством итераций равной 1000. По результатам компьютерных симуляций была построена зависимость значений функции ценности от поколений, которая приведена на рисунке 4.

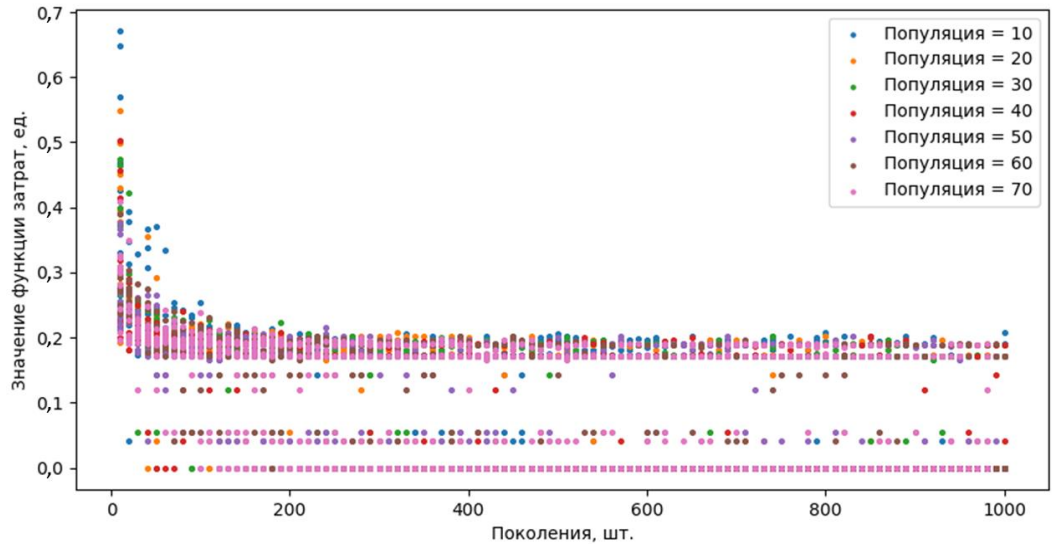

Рис. 4 – Зависимость значения функции затрат от количества поколений для различны значений популяции

Fig. 4 – Dependence of the cost function value on the number of generations for different population values

Как можно видеть, с ростом размера популяции увеличивается количество успешных решений, на графике это достижение точки 0. Также стоит отметить, что в диапазоне популяции от 10 до 20 практически отсутствуют решения, а большое количество итераций не способствует поиску решений, так как формируются локальные минимумы, которые достаточно сложно покинуть за счет невозможности краткосрочно пренебречь функцией ценности, чтобы в будущем получить более оптимально решение.

В результате полученных данных, был построен график 5 зависимости количества успешных найденных параметров от диапазонов поколений $D_1, D_2, D_3, D_4$ и $D_5$ такими, что:
$D_1 = \{x \in R: 0 < x \leq 200\};$   $D_2 = \{x \in R: 200 < x \leq 400\};$ $D_3 = \{x \in R: 400 < x \leq 600\};$ $D_4 = \{x \in R: 600 < x \leq 800\}; D_5 = \{x \in R: 800 < x \leq 1000\}.$

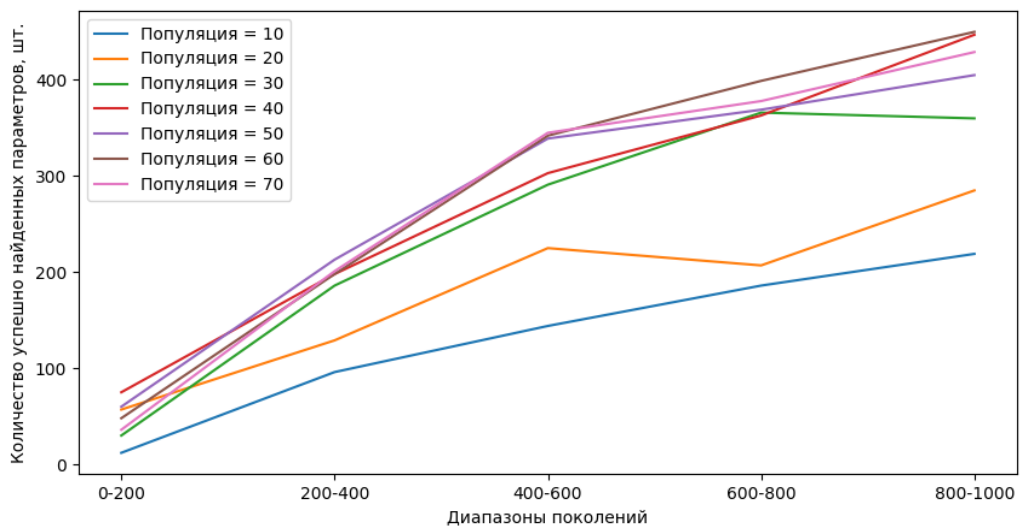

Рис.5 – Зависимость количества успешно найденных параметров от диапазонов поколений

Fig.5 – Dependence of the number of successfully found parameters on the generation ranges

Таким образом можно сделать вывод, что зависимость нелинейная, которая имеет достаточно быстрый рост от первого ко второму диапазону, а также от второго к третьему. На первом диапазоне практически нет успешных решений. После третьего диапазона рост замедляется от предыдущего на 29.4% и 20% соответственно. Также стоит отметить, что после популяции в 30 единиц смещение графика достаточное небольшое, в таком случае можно говорить о том, что использование популяции с более высокими значениями будет более ресурсоемким, но прирост эффективности от этого будем много меньше.

Теперь проведем сравнение предложенного решения с методом синтеза функции Ляпунова, основанным на контраргументах и нейронной сети [15]. В качестве сравнения была взята планарная динамическая система, описываемая следующей моделью:

$$\begin{cases} \dot{x} = -x + xy \\ \dot{y} = -y \end{cases} \tag{15}$$

Сравнению подлежала область $\Omega$ и временные затраты на поиск функции Ляпунова. Результаты сравнительного анализа приведены в таблице 3.

Таблица 3 – Результаты сравнения нашего метода с предложенным в работе [15], выраженные временем (t, сек), в зависимости от области $\Omega$

| Метод (алгоритм) | Область $\Omega$ | | | | | |
|---|---|---|---|---|---|---|
| | 10 | 20 | 50 | 100 | 200 | 500 |
| Метод [15] | 0,06 | 0,14 | 0,11 | 1,90 | 23,10 | 12,01 |
| Алгоритм 1 | 0,18 | 0,12 | 0,21 | 0,91 | 2,19 | 7,92 |

Как видно из таблицы 3, предложенный метод обеспечивает более эффективный способ поиска функции Ляпунова с точки зрения временных затрат. Также из таблицы 3 можно сделать вывод, что предложенный алгоритм синтезировал функцию Ляпунова быстрее для области 20, 100, 200 и 500. Дополнительно следует отметить, что в данной работе рассматривается квадратная область, а в [15] окружность. Таким образом в данной работе осуществляется проверка условия теоремы Ляпунова для большей области.

**Оценка сходимости генетического алгоритма**

Генетический алгоритм является стохастическим алгоритмом поиска, откуда следует, что нельзя гарантировать поиск оптимального решения за фиксированное количество итераций, но возможно исследовать сходимость алгоритма по вероятности. Иными словами, можно гарантировать, что будет найдено хотя бы одно оптимальное решение с вероятностью не менее $p_{conv}$ за фиксированное число итераций $\tau$ [18]:

$$\tau(p_{conv}) = INT\left[\frac{\ln(1-p_{conv})}{n \cdot ln\left(1-\min\left[(1-\mu)^{\gamma-1}\left(\frac{\mu}{K-1}\right),\left(\frac{\mu}{K-1}\right)^{\gamma}\right]\right)}\right], \quad (16)$$

где $\mu$ – вероятность мутации; $\gamma$ – количество оптимизированных параметров; $K$ – количество различных значений, которые могут принимать параметры; $n$ – размер популяции; $INT$ – инструкция для перевода значения в целочисленное.

На рисунке 6 представлена зависимость вероятности получения как минимум одной потенциальной функции в популяции, удовлетворяющей теореме Ляпунова, и необходимого для этого числа итераций генетического алгоритма. Предложенный в работе алгоритм гарантирует нахождение потенциальной функции Ляпунова для примера (11)-(12) вероятностью не менее 80% за 314, 1357 и 4186 итераций для вероятности мутации параметров 20%, 17% и 15% соответственно.

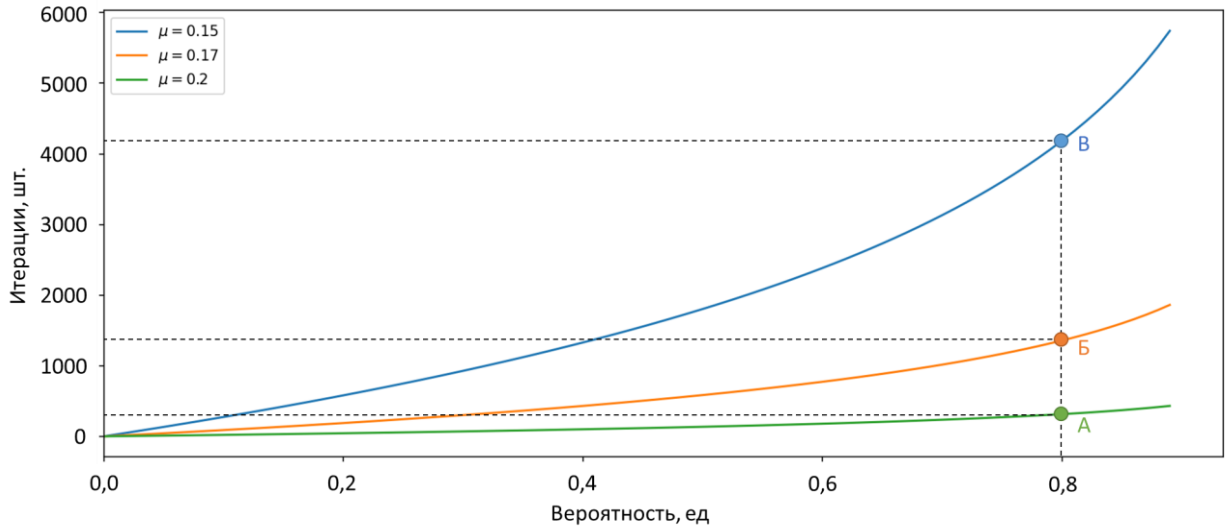

Рис. 6 – Зависимость вероятности получения как минимум одной потенциальной функции в популяции, удовлетворяющей теореме Ляпунова, и необходимого для этого числа итераций генетического алгоритма

Fig. 6 – The dependence of the probability of obtaining at least one potential function in the population satisfying the Lyapunov theorem and the number of iterations of the genetic algorithm required for this

Также для теоретического анализа генетического алгоритма используется [19] теорема о схемах Холланда [20]. Суть теоремы заключается в том, что пригодность целевой функции гарантированно улучшается по мере увеличения числа итераций, другими словами, генетический алгоритм на каждом шаге приближается к истинному минимуму целевой функции. Математически это можно выразить следующим образом:

$$N(h, t+1) \geq N(h,t)\frac{f(h,t)}{f(t)}\left(1 - \frac{\sigma(h)}{l-1}p_c - o(h)p_m\right), \quad (17)$$

где $N(h,t)$ – количество примеров схемы h на шаге t; $N(h, t+1)$ – количество примеров схемы h на шаге t+1; $f(h,t)$ – функция пригодности схемы на шаге t; $f(t)$ – среднее значение функции пригодности по всей популяции на шаге t; $\sigma(h)$ – определяющая длина схемы h; $o(h)$ – порядок схемы h; $p_c$ – вероятность скрещивания; $p_m$ – вероятность мутации; $l$ – длина последовательности.

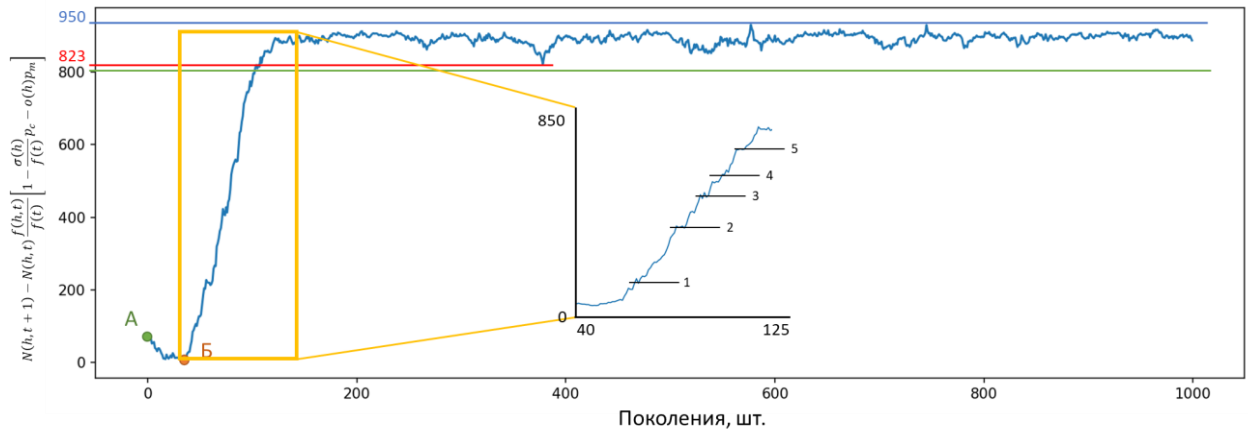

Рис. 7 – Иллюстрация теоремы схем для предлагаемого решения
Fig. 7 - Illustration of the circuit theorem for the proposed solution

На рисунке 7 показана иллюстрация неравенства (17) для предлагаемого в работе решения. На участке АБ отсутствовали коэффициенты, которые вносили наибольший вклад в функцию пригодности, поэтому можно наблюдать спад, который в то же время не пересекает отметку нуля. Далее идет экспоненциальный рост и попадание в коридор со значениями от 807,5 до 950. Данный коридор обусловлен вероятностью функции мутации ГА, которая составляет 15%. Также можно наблюдать небольшие колебания в точках 1–5, которые обусловлены попаданием алгоритма в локальные минимумы, где большую составляющую играет функция мутации, на этих участках также отсутствует колебания больше 15%. Таким образом данный график иллюстрирует выполнение оценки (17) и подтверждает, что известные аналитические результаты о характере сходимости генетических алгоритмов выполнены для предложенного алгоритма 1.

**Заключение**

В статье был предложен автоматический подход для синтеза функции Ляпунова для нелинейных динамических систем путем разложения потенциальной функции в ряд Тейлора, постоянные коэффициенты которого предполагались неизвестными. Поиск этих коэффициентов осуществлялся с использованием генетического алгоритма. Также в работе был предложен метод определения функции затрат для оценки эффективности поиска параметров генетическим алгоритмом. Предложенные решения были апробированы на модели математического маятника с различными диапазонами параметров и областью $\Omega$. Проведен сравнительный анализ предложенного подхода с альтернативным решением [13], использующим для синтеза функции Ляпунова технологии машинного обучения. Предложенный в рамках этой статьи метод показал большую эффективность с точки зрения временных затрат. Предложенный метод отличается более общим подходом, поскольку рассматривается разложение потенциальной функции в ряд Тейлора с неизвестными коэффициентами, вместо использования контрпримеров или шаблонных функций Ляпунова. Также стоит отметить, что данный метод можно применять для синтеза сигнала управления нелинейными систем, в основе которого используется функция Ляпунова, поиск которой происходит через предложенный метод.

Наряду с этим функция затрат требует учет дополнительных параметров системы, а генетический алгоритм имеет тенденцию попадания в локальный минимум. Также стоит отметить увеличение времени поиска функции Ляпунова с увеличением степени разложения ряда, что является достаточно логичным, так как на это уходит больше вычислительных ресурсов.

## Литература